\documentclass[letterpaper,spanish]{revtex4}
\usepackage[T1]{fontenc}
\usepackage[latin1]{inputenc}
\usepackage{amsmath}
\usepackage{graphicx}
\usepackage{amssymb}

\makeatletter
\usepackage{graphicx}
\usepackage{amsfonts}
\usepackage{babel}
\deactivatetilden
\makeatother
\begin{document}

\title{Traversable Wormholes Construction in (2+1) Gravity}

\author{Eduard Alexis Larrañaga Rubio}

\email{eduardalexis@gmail.com}

\address{National 0University of Colombia}

\address{National Astronomical Observatory (OAN)}

\begin{abstract}
Wormholes have been always an interesting object in gravity theories.
In this paper we make a little review of the principal properties
of these objects and the exotic matter they need to exist. Then, we
obtain two specific solutions in the formalism of (2+1)-dimensional
gravity with negative cosmological constant. The obtained geometries
correspond to traversable wormholes with an exterior geometry correspondient
to the well known BTZ black hole solution. We also discuss the distribution
of exotic matter that these wormholes need. 
\end{abstract}
\maketitle

\section{Introduction}

20 years ago Morris-Thorne\cite{morris} propose the traversable wormholes
as a teaching tool for General Relativity. In 1995, the well known
book of Visser collects all the work done in this area on the last
century. Today, wormholes are one of the most studied solutions of
Einstein's field equations because of thei characetristics and theis
relation with the existence of a special type of matter that would
violate the energy conditions, now called \emph{exotic matter}. \\

On the other side, the ($2+1$) dimensional gravity is a covariant
theory of spacetime geometry that has a great simplicity when compared
with General Relativity, and this made of it a good theory to study
some quantum aspects of gravyty The most interesting solution of this
theory, is the black hole metric found by Banados, Teitelboim and
Zanelli (BTZ Blak Hole)\cite{BTZ1,BTZ2} in a universe with a negative
cosmological constant. \\
Some years ago, some authors are interested on wormholes in ($2+1$)
dimensional gravity. Delgaty et. al. \cite{delgaty} made an analysis
of the characteristic of one specific wormhole in universes with cosmological
constant. Aminneborg et. al. \cite{Amin} compares the properties
of wormholes with the characteristic of black holes while Kim et.
al. \cite{Kim} give two specific solutions taking ($2+1$) dimensional
gravity with a dilatonic field.\\

In this paper we give two specific wormhole solutions in ($2+1$)
gravity obtained joining two spacetime manifolds following the work
of Lemos et. al. \cite{Lemos}. The exterior of this wormholes correspond
to the BTZ metric withut electric charge and without angular momentum.
We also show the matter distribution tath is needed to mantain the
solutions.

\section{Structure Equations for the Wormhole}

\subsection{Traversable Wormholes Properties}

In order to restrain the possible solutions to traversable wormholes
we must impose some conditions \cite{morris}: 

\begin{enumerate}
\item The metric must be a solution of the field equations at every spacetime
point. 
\item Metric must have spherical symmetry and must be static (i.e. time
independient).
\item Solution must have a {}``throat'' that connects two spacetime regions.
In the ($2+1$) case, the exterior spacetime must correspond to BTZ
solution. 
\item Metric must not have event horizons (it would prevent two way travel)
\item Tidal forces must be small or null (we want to be possible to travel
into the wormhole )
\item The time needed to cross the wormhole must be reasonable. 
\end{enumerate}
These conditions will be imposed in order to obain the specific solutions. 

\bigskip

The wormhole metric must be spherically symmetric and time independient.
Thus, in usual spherical coordinates $\left(t,r,\varphi\right)$,
we have the general metric \begin{equation}
ds^{2}=-e^{2\Phi\left(r\right)}dt^{2}+\frac{1}{1-\frac{b\left(r\right)}{r}}dr^{2}+r^{2}d\varphi^{2},\label{metrica}\end{equation}

where $\Phi\left(r\right)$ , $b\left(r\right)$ are arbitrary functions
of the radial coordinate a $r$, that will be restrained by the imposed
conditions. The function $\Phi\left(r\right)$ is known as \textit{{}``}\emph{redshift
function''}, while $b\left(r\right)$ is the \emph{{}``shape function''}. 

The field equations with a cosmological constant can be written as

\begin{equation}
G_{\mu v}+\Lambda g_{\mu v}=T_{\mu v}\label{eccampo}\end{equation}

where $g_{\mu v}$ is the metric, $\Lambda$ is the cosmological constant,
$T_{\mu v}$ is the stress-energy tensor and $G_{\mu v}$ es is the
Einstein tensor defined by \begin{equation}
G_{\mu v}=R_{\mu v}-\frac{1}{2}g_{\mu v}R\end{equation}

with $R_{\mu v}$ the Ricci tensor. Remember that greek indice take
the values $0,1,2$.

\bigskip

Using the metric (\ref{metrica}), the Einstein tensor has components 

\bigskip\begin{align}
G_{tt} & =\frac{1}{2r^{3}}e^{2\Phi}\left[-b+rb^{\prime}\right]\label{einstein1}\\
G_{rr} & =\frac{\Phi^{\prime}}{r}\nonumber \\
G_{\varphi\varphi} & =\frac{1}{2}\left[\Phi^{\prime}\left(b-rb^{\prime}\right)+2r\left(r-b\right)\left(\left(\Phi^{\prime}\right)^{2}+\Phi^{\prime\prime}\right)\right],\nonumber \end{align}

where primes represent derivative with respect to the radial coordinate
$r$.

\bigskip

\subsection{Change of Basis}

The Einstein tensor obtained above is defined using the $\left(\mathbf{e}_{t},\mathbf{e}_{r},\mathbf{e}_{\varphi}\right)$
triad associated with the coordinates $t,r,\varphi$. However, we
can choose any coordinate system and, in this case, is useful to consider
a set of orthonormal vectors as basis. These correspond to the reference
system of an observer that remains at rest in the $\left(t,r,\varphi\right)$
system; i.e. with $r,\varphi$ constant.

We will denote this triad by $\left(\mathbf{e}_{\widehat{t}},\mathbf{e}_{\widehat{r}},\mathbf{e}_{\widehat{\varphi}}\right)$;
and its relation with the original $\left(\mathbf{e}_{t},\mathbf{e}_{r},\mathbf{e}_{\varphi}\right)$
is\begin{align}
\mathbf{e}_{\widehat{t}} & =e^{-\Phi}\mathbf{e}_{t}\\
\mathbf{e}_{\widehat{r}} & =\left(1-\frac{b}{r}\right)^{1/2}\mathbf{e}_{r}\\
\mathbf{e}_{\widehat{\varphi}} & =\frac{1}{r}\mathbf{e}_{\varphi}\end{align}

It is important to note that in the new system, the metric tensor
is \begin{equation}
g_{\alpha\beta}=\mathbf{e}_{\widehat{\alpha}}\cdot\mathbf{e}_{\widehat{\beta}}=\eta_{\widehat{\alpha}\widehat{\beta}}=\left[\begin{array}{ccc}
-1 & 0 & 0\\
0 & 1 & 0\\
0 & 0 & 1\end{array}\right]\end{equation}

and all the tensor will change their components. For example, the
field equation will take the form \begin{equation}
G_{\widehat{\mu}\widehat{v}}+\Lambda\eta_{\widehat{\mu}\widehat{v}}=T_{\widehat{\mu}\widehat{v}}.\label{eccampo2}\end{equation}

\bigskip

where the new Einstein tensor is now given by \begin{align}
G_{\widehat{t}\widehat{t}} & =\frac{1}{2r^{3}}\left[b^{\prime}r-b\right]\label{teinstein1}\\
G_{\widehat{r}\widehat{r}} & =\left(1-\frac{b}{r}\right)\frac{\Phi^{\prime}}{r}\\
G_{\widehat{\varphi}\widehat{\varphi}} & =\frac{1}{2r^{2}}\left[\Phi^{\prime}\left(b-rb^{\prime}\right)+2r\left(r-b\right)\left(\left(\Phi^{\prime}\right)^{2}+\Phi^{\prime\prime}\right)\right]\label{teinstein3}\end{align}

\bigskip

\section{Stress- Energy Tensor }

In order to obtain a traversable wormhole, we need a non null Stress-Energy
tensor. Since the field equations (\ref{eccampo2}) tell us that the
stress-energy tensor is proportional to Einstein tensor, they must
have the same algebraic structure, i. e. that the non zero components
of $T_{\widehat{\mu}\widehat{v}}$ must be $T_{\widehat{t}\widehat{t}},T_{\widehat{r}\widehat{r}}$
and $T_{\widehat{\varphi}\widehat{\varphi}}$.

In the orthonormal basis $\left(\mathbf{e}_{\widehat{t}},\mathbf{e}_{\widehat{r}},\mathbf{e}_{\widehat{\varphi}}\right)$
related with the inertial reference system of an static observer,
the components of the stress-energy tensor take an immediate interpretation,
\begin{align}
T_{\widehat{t}\widehat{t}} & =\rho\left(r\right)\label{tmomento}\\
T_{\widehat{r}\widehat{r}} & =-\tau\left(r\right)\nonumber \\
T_{\widehat{\varphi}\widehat{\varphi}} & =p\left(r\right),\nonumber \end{align}

where $\rho\left(r\right)$ is the mass-energy density, $\tau\left(r\right)$
is the radial tension per unit area (i.e. the negative of the radial
pressure, $\tau\left(r\right)=-p_{r}\left(r\right)$) and $p\left(r\right)$
is the tangential pressure.

Sometimes its interesting to write the cosmological constant term
in the field equations (\ref{eccampo2}) as

\begin{equation}
T_{\widehat{\mu}\widehat{v}}^{\left(vac\right)}=-\Lambda\eta_{\widehat{\mu}\widehat{v}}=\left[\begin{array}{ccc}
\Lambda & 0 & 0\\
0 & -\Lambda & 0\\
0 & 0 & -\Lambda\end{array}\right],\end{equation}

and then the field equations will be\begin{equation}
G_{\widehat{\mu}\widehat{v}}=\left(T_{\widehat{\mu}\widehat{v}}+T_{\widehat{\mu}\widehat{v}}^{\left(vac\right)}\right)\end{equation}
\begin{equation}
G_{\widehat{\mu}\widehat{v}}=\overline{T}_{\widehat{\mu}\widehat{v}},\end{equation}

where $\,\,\overline{T}_{\widehat{\mu}\widehat{v}}=T_{\widehat{\mu}\widehat{v}}+T_{\widehat{\mu}\widehat{v}}^{\left(vac\right)}$
is the total stress-energy tensor. Therefore, we can define the functions

$\overline{\rho}\left(r\right),\overline{\tau}\left(r\right)$ and
$\overline{p}\left(r\right)$ by\begin{align}
\overline{\rho}\left(r\right) & =\rho\left(r\right)+\Lambda\\
\overline{\tau}\left(r\right) & =\tau\left(r\right)+\Lambda\nonumber \\
\overline{p}\left(r\right) & =p\left(r\right)-\Lambda\nonumber \end{align}

\section{Solving the Field equations}

Using the Einstein tensor given by (\ref{teinstein1}-\ref{teinstein3})
and the stres-energy tensor (\ref{tmomento}), we can obtain the field
equations,\begin{equation}
\rho\left(r\right)=\frac{1}{2r^{3}}\left[b^{\prime}r-b\right]-\Lambda\label{aux1}\end{equation}
\begin{equation}
\tau\left(r\right)=-\left(1-\frac{b}{r}\right)\frac{\Phi^{\prime}}{r}-\Lambda\label{aux2}\end{equation}
\begin{equation}
p\left(r\right)=\frac{1}{2r^{2}}\left[\Phi^{\prime}\left(b-rb^{\prime}\right)+2r\left(r-b\right)\left(\left(\Phi^{\prime}\right)^{2}+\Phi^{\prime\prime}\right)\right]+\Lambda.\label{aux3}\end{equation}

Taking the derivative of equation (\ref{aux2}) with respect to $r$
we have\begin{align}
\tau^{\prime}\left(r\right) & =-\left(1-\frac{b}{r}\right)\frac{\Phi^{\prime\prime}}{r}+\left(1-\frac{b}{r}\right)\frac{\Phi^{\prime}}{r^{2}}+\frac{b^{\prime}r-b}{r^{3}}\Phi^{\prime}.\end{align}

Using equations (\ref{aux1}-\ref{aux3}) to eliminate $b^{\prime}$
and $\Phi^{\prime\prime}$ we obtain

\begin{equation}
\tau^{\prime}\left(r\right)=\left(\rho-\tau\right)\Phi^{\prime}-\frac{p+\tau}{r}\label{aux4}\end{equation}

Equations (\ref{aux1}), (\ref{aux3}) and (\ref{aux4}) are three
differential equations that correlate the five unknown functions $b,\Phi,\rho,\tau$
and $p$.

Now, the usual way to solve these equations is to assume a specific
kind of matter and energy. The corresponding state equation gives
a relationship between the tension and the density function, $\tau\left(\rho\right)$,
and between the pressure and density, $p\left(\rho\right)$. Therefore,
we got five equations for five unknown functions, and we can find
the form of the spacetime manifold, i.e. we can obtain the functions
$b\left(r\right)$ and $\Phi\left(r\right)$.

\bigskip

For wormholes, we proceed in a different way: we impose the conditions
on the geometry of the spacetime manifold (i. e. we impose the functions
$b\left(r\right)$ and $\Phi\left(r\right)$), and using the field
equations we obtain the needed matter-energy distribution for that
geometry.

\bigskip

\subsection{Geometry of the Wormhole}

The wormhole metric given by (\ref{metrica}), considered for a fixed
time $t$ is\begin{equation}
ds^{2}=\frac{dr^{2}}{1-\frac{b}{r}}+r^{2}d\varphi^{2}.\label{corte}\end{equation}

If we make an embedding of this metric in the three-dimensional euclidean
space with cylindrical coordinates,\begin{equation}
ds^{2}=dz^{2}+dr^{2}+r^{2}d\varphi^{2},\end{equation}

we obtain the equation for the embedding surface, \begin{equation}
\frac{dz}{dr}=\pm\left(\frac{r}{b}-1\right)^{-1/2}.\label{aux5}\end{equation}

To obtain a wormhole geometry, the solution must have a minimum radius
called {}``throat'', $r=b\left(r\right)=r_{m}$. At the throat the
embedded surface is vertical, i.e. $\frac{dz}{dr}\rightarrow\infty$.
On the other hand, far from the mouth of the wormhole, the space is
asymptotically flat, i.e. $\frac{dz}{dr}\rightarrow0$. 

Since the wormhole metric must be connected smoothly with the exterior
spacetime, the throat must flare out. this condition can be written
in terms of the embedding function as

\begin{equation}
\frac{d^{2}r}{dz^{2}}>0.\end{equation}

\bigskip Using equation (\ref{aux5}) we have \begin{equation}
\frac{dr}{dz}=\pm\left(\frac{r}{b}-1\right)^{1/2},\end{equation}

and differentitating with respect to $z$ we obtain\begin{equation}
\frac{d^{2}r}{dz^{2}}=\pm\frac{1}{2}\left(\frac{b-rb^{\prime}}{b^{2}}\right).\end{equation}

Therefore, the flare out condition is \begin{equation}
\frac{d^{2}r}{dz^{2}}=\frac{b-rb^{\prime}}{2b^{2}}>0\text{ \,\,\,\, at the throath or near }\end{equation}

\bigskip

Now, in order to assure that the wormhole permits inside and outside
travel, we need that there will be no event horizon. For static metrics,
horizons correspond to non singular surfaces at which \[
g_{tt}=-e^{2\Phi}\rightarrow0.\]

Then, in order to assure traversability, we need that the function
$\Phi\left(r\right)$ \textit{be finite at every point.}

\section{Properties of the Stress-Energy Tensor }

If we define the adimensional function \begin{equation}
\varsigma=\frac{\tau-\rho}{\left|\rho\right|},\end{equation}

the wormhole field equations (\ref{aux1}) and (\ref{aux2}) give
\begin{equation}
\varsigma=\frac{\tau-\rho}{\left|\rho\right|}=\frac{-2r^{2}\left(1-\frac{b}{r}\right)\Phi^{\prime}-\left(b^{\prime}r-b\right)}{\left|b^{\prime}r-b-2r^{3}\Lambda\right|}.\label{aux22}\end{equation}

To obtain a wormhole, we must demand that the inside metric joins
soomthly with the outside metric ( that corresponds to BTZ metrc),
and then we impose the flare out condition described above. This condition
is given by \begin{equation}
\frac{d^{2}r}{dz^{2}}=\frac{b-rb^{\prime}}{2b^{2}}>0\text{ \,\,\,\, at the throath or near it.}\end{equation}

Then, equation (\ref{aux22}) is\begin{equation}
\varsigma=\frac{\tau-\rho}{\left|\rho\right|}=\frac{2b^{2}}{\left|2b^{2}\frac{d^{2}r}{dz^{2}}+2r^{3}\Lambda\right|}\frac{d^{2}r}{dz^{2}}-2\left(1-\frac{b}{r}\right)\frac{r^{2}\Phi^{\prime}}{\left|2b^{2}\frac{d^{2}r}{dz^{2}}+2r^{3}\Lambda\right|}.\end{equation}

Near the throat we have $\left(1-\frac{b}{r}\right)\Phi^{\prime}\longrightarrow0$.
Therefore the flare out condition implicates \begin{equation}
\varsigma_{m}=\frac{\tau_{m}-\rho_{m}}{\left|\rho_{m}\right|}>0,\label{aux23}\end{equation}

where de index $_{m}$ indicates that we are evaluating at the thorath
or near it. 

The condition $\tau_{m}>\rho_{m}$ is imposed by (\ref{aux23}) and
any material that satisfies the property $\left(\tau_{m}>\rho_{m}>0\right)$
is called {}``exotic'' and will violate the \emph{energy conditions}\cite{larranaga}. 

\bigskip

\section{Construction of Wormholes}

In order to construct the wormholes we use the equations that relate
$b,\Phi,\rho,\tau$ y $p$,\begin{align}
\rho\left(r\right) & =\frac{1}{2r^{3}}\left[b^{\prime}r-b\right]-\Lambda\label{aux7}\\
\tau\left(r\right) & =-\left(1-\frac{b}{r}\right)\frac{\Phi^{\prime}}{r}-\Lambda\nonumber \\
p\left(r\right) & =\frac{1}{2r^{2}}\left[\Phi^{\prime}\left(b-rb^{\prime}\right)+2r\left(r-b\right)\left(\left(\Phi^{\prime}\right)^{2}+\Phi^{\prime\prime}\right)\right]+\Lambda\nonumber \\
\tau^{\prime}\left(r\right) & =\left(\rho-\tau\right)\Phi^{\prime}-\frac{p+\tau}{r}.\nonumber \end{align}

Since we work with a cosmológical constant we will distinguish between
the inside solution( i.e. $r<a$, con $\Lambda_{int}$) and the outside
solution (i.e. $r>a$, con $\Lambda_{ext}$).

\subsection{Interior Solution}

The interior solution must have the wormhole form \begin{equation}
ds^{2}=-e^{2\Phi^{int}\left(r\right)}c^{2}dt^{2}+\frac{1}{1-\frac{b^{int}\left(r\right)}{r}}dr^{2}+r^{2}d\varphi^{2}.\label{metricainterior}\end{equation}

\bigskip

To find explicitly the functions $\Phi_{int}\left(r\right)$ and $b_{int}\left(r\right)$
inside ($r<a$), we will use $\Lambda_{int}$ in th equations (\ref{aux7}).
(We will not use the index $^{int}$ in the functions $\Phi$ y $b$)\begin{equation}
\rho\left(r\right)=\frac{1}{2r^{3}}\left[b^{\prime}r-b\right]-\Lambda_{int}\label{aux14}\end{equation}
\begin{equation}
\tau\left(r\right)=-\left(1-\frac{b}{r}\right)\frac{\Phi^{\prime}}{r}-\Lambda_{int}\label{aux13}\end{equation}
\begin{equation}
p\left(r\right)=\frac{1}{2r^{2}}\left[\Phi^{\prime}\left(b-rb^{\prime}\right)+2r\left(r-b\right)\left(\left(\Phi^{\prime}\right)^{2}+\Phi^{\prime\prime}\right)\right]+\Lambda_{int}\label{aux15}\end{equation}

In the equation for the tension, we see that at the throat ($b\left(r_{m}\right)=r_{m}$),
we have 

\bigskip\begin{equation}
\tau\left(r_{m}\right)=-\left(1-\frac{r_{m}}{r_{m}}\right)\frac{\Phi^{\prime}}{r_{m}}-\Lambda_{int}\end{equation}
\begin{equation}
\tau\left(r_{m}\right)=-\Lambda_{int}.\end{equation}

i. e. that the radial tension at the throat is positive for holes
with $\Lambda_{int}<0$ and is negative (i. e. a pressure) for holes
with $\Lambda_{int}>0$. On the other side, it is interesting to note
that the total radial tension at the throat is zero , \begin{equation}
\overline{\tau}\left(r_{m}\right)=\tau\left(r_{m}\right)+\Lambda_{int}=0.\end{equation}

\subsection{Exterior Solution }

In the exterior of the wormhole ($r>a$) we consider a vaccum spaetime
geometry, i. e. a null stress-energy tensor $T_{\widehat{\mu}\widehat{v}}=0$.
This means 

\begin{equation}
\rho\left(r\right)=\tau\left(r\right)=p\left(r\right)=0.\end{equation}

However, we may have a non null exterior cosmological constant $\Lambda_{ext}$
. Equations (\ref{aux7}) will be now\begin{align}
0 & =\frac{1}{2r^{3}}\left[b^{\prime}r-b\right]-\Lambda_{ext}\\
0 & =-\left(1-\frac{b}{r}\right)\frac{\Phi^{\prime}}{r}-\Lambda_{ext}\nonumber \\
0 & =\frac{1}{2r^{2}}\left[\Phi^{\prime}\left(b-rb^{\prime}\right)+2r\left(r-b\right)\left(\left(\Phi^{\prime}\right)^{2}+\Phi^{\prime\prime}\right)\right]+\Lambda_{ext}.\nonumber \end{align}

Solving this equations we obtain the exterior solution\cite{larranaga},\begin{equation}
ds^{2}=-\left(-M-\Lambda_{ext}r^{2}\right)dt^{2}+\frac{dr^{2}}{\left(-M-\Lambda_{ext}r^{2}\right)}+r^{2}d\varphi^{2}.\label{metricaexterior}\end{equation}

If the cosmológical constant is negative $\Lambda_{ext}<0$, we can
write it, following Banados et. al.\cite{BTZ1,BTZ3}, as

\begin{equation}
\Lambda_{ext}=-\frac{1}{l^{2}},\end{equation}

and the exterior solution gives the usual BTZ black hole metric,

\begin{equation}
ds^{2}=-\left(-M+\frac{r^{2}}{l^{2}}\right)dt^{2}+\frac{dr^{2}}{\left(-M+\frac{r^{2}}{l^{2}}\right)}+r^{2}d\varphi^{2}.\end{equation}

Note that this solution have singularities at the radii

\begin{equation}
r_{\pm}=\pm\sqrt{M}l.\end{equation}

The outside singularity $r_{+}$ corresponds to the event horizon
for the black hole and in order to satisfy the traversability conditions,
we must impose $a>r_{+}$.

\bigskip

\subsection{Junction Conditions}

In order to join the interior and exterior metrics we consider the
boundary surface $S$ that connects them. The first condition is that
the metric must be continuous at $S$, i. e. $\left.g_{\mu v}^{int}\right|_{S}=\left.g_{\mu v}^{ext}\right|_{S}$. 

However, this condition is not enough to make the junction. The Darmois-Israel
formalism impose the continuity of the second fundamental form (extrinsic
curvature) at the surface $S$. But, when the spacetime is spherically
symmetric, the second condition can be done directly with the field
equations.  \\

With these conditions, we will find the stress-energy density at surface
$S$ needed to made the junction between the exterior and interior
regions. When there is no stress-energy terms at $S$, we say that
this is a \emph{boundary surface}, while when we have som stress-energy
terms we call it a \textit{thin-shell}.

\subsubsection{Continuity of the metric }

Since both the inside and outside metrics are spherically symmetric,
then the continuity of the metric condition $\left.g_{\mu v}^{int}\right|_{S}=\left.g_{\mu v}^{ext}\right|_{S}$
is immediate for the $g_{\varphi\varphi}$component. For the \emph{t}
and \emph{r} components \emph{}we impose\begin{align}
\left.g_{tt}^{int}\right|_{r=a} & =\left.g_{tt}^{ext}\right|_{r=a}\\
\left.g_{rr}^{int}\right|_{r=a} & =\left.g_{rr}^{ext}\right|_{r=a}.\nonumber \end{align}

Using equations (\ref{metricainterior}) and (\ref{metricaexterior}),
the continuity conditions are \begin{equation}
e^{2\Phi\left(a\right)}=-M+\frac{a^{2}}{l^{2}}\label{aux11}\end{equation}
\begin{equation}
1-\frac{b\left(a\right)}{a}=-M+\frac{a^{2}}{l^{2}},\label{aux8}\end{equation}

that can be written as \begin{equation}
\Phi\left(a\right)=\frac{1}{2}\ln\left(-M+\frac{a^{2}}{l^{2}}\right)\end{equation}
\begin{equation}
b\left(a\right)=\left(1+M\right)a-\frac{a^{3}}{l^{2}}.\label{aux12}\end{equation}

Last equation let us obtain an expresion for the wormhole mass,\begin{equation}
M=\frac{b\left(a\right)}{a}+\frac{a^{2}}{l^{2}}-1.\label{masaagujero}\end{equation}

\subsubsection{Field Equations}

To complete the junction of exterior and interior metrics we will
use the field equations (\ref{eccampo}). We also suppose that static
observers inside will feel null tidal forces, i.e. e $\Phi^{int}=$constant,
and therefore we have $\Phi^{\prime int}=0$.

If we have a thin-shell, the components of the stress-energy tensor
is non zero at the surface $S$ and we an write them as proportional
to the Dirac's delta function, \begin{equation}
T_{\widehat{\mu}\widehat{v}}=t_{\widehat{\mu}\widehat{v}}\delta\left(\widehat{r}-\widehat{a}\right),\end{equation}
 where $\widehat{r}=\sqrt{g_{rr}}r$ is the proper radial distance
measured inside the thin-shell. To obtain the components $t_{\widehat{\mu}\widehat{v}}$
we must use \begin{equation}
\int_{int}^{ext}G_{\widehat{\mu}\widehat{v}}d\widehat{r}=\int_{int}^{ext}t_{\widehat{\mu}\widehat{v}}\delta\left(\widehat{r}-\widehat{a}\right)d\widehat{r},\end{equation}

where $\int_{int}^{ext}$ is an infinitesiaml integral along the thin-shell.
Using delta function property \begin{equation}
\int g\left(x\right)\delta\left(x-x_{o}\right)dx=g\left(x_{o}\right),\end{equation}

we have\begin{equation}
t_{\widehat{\mu}\widehat{v}}=\int_{int}^{ext}G_{\widehat{\mu}\widehat{v}}d\widehat{r}.\end{equation}

\paragraph{Surface Pressure}

Now we will consider the surface energy density and surface tangential
presure terms. From (\ref{teinstein1}) we can see that the $G_{\widehat{t}\widehat{t}}$
component depends only on the first derivatives of the metric. Therefore,
the surface energy density is \begin{equation}
\Sigma=t_{\widehat{t}\widehat{t}}=\int_{int}^{ext}G_{\widehat{t}\widehat{t}}d\widehat{r}.\end{equation}

When making the integration we will only obtain functions of the metric,
and they are continuous because of the continuity condition for the
metric. Since the integral is evaluated in the interior and exterior
regions, the final integral vanishes. Hence, we have \begin{equation}
\Sigma=0.\end{equation}

\bigskip

On the other side, from equations (\ref{teinstein1}) we see taht
the $G_{\widehat{\varphi}\widehat{\varphi}}$ component has terms
tht depend on the first derivatives of the metric, and they will not
contribute to the total integral. However, there is also a term with
the form $\left(1-\frac{b}{r}\right)\Phi^{\prime\prime}$. This term
doesn't cancel out, and therefore, the surface tangential pressure
can be written as \begin{equation}
\mathcal{P}=t_{\widehat{\varphi}\widehat{\varphi}}=\int_{int}^{ext}G_{\widehat{\varphi}\widehat{\varphi}}d\widehat{r}\end{equation}
\begin{equation}
\mathcal{P}=\left[\sqrt{1-\frac{b\left(a\right)}{a}}\left.\Phi^{\prime}\right|_{int}^{ext}\right].\end{equation}

Since we assume that a static internal observer does not feel any
tidal force, we have $\Phi^{\prime int}=0$. We also have \begin{align}
\Phi^{\prime ext} & =\frac{a}{l^{2}}\left(-M+\frac{a^{2}}{l^{2}}\right)^{-1}.\end{align}

Using (\ref{aux8}) we obtain\begin{equation}
\Phi^{\prime ext}=\frac{\frac{a}{l^{2}}}{\left(1-\frac{b\left(a\right)}{a}\right)},\label{aux9}\end{equation}

and then, the surface tangential pressure is

\bigskip\begin{equation}
\mathcal{P}=\frac{\frac{a}{l^{2}}}{\sqrt{1-\frac{b\left(a\right)}{a}}}\label{aux10}\end{equation}

\begin{equation}
\mathcal{P}=\frac{\frac{a}{l^{2}}}{\sqrt{-M+\frac{a^{2}}{l^{2}}}}.\end{equation}

Note that in this case, the tangential pressure is always positive,
under the condition

\begin{equation}
a^{2}\geq Ml^{2},\end{equation}
that corresponds to say that the wormhole's mouth is outside the event
horizon correspondient to the exterior BTZ metric. \bigskip

\paragraph{Radial Pressure}

The radial component of the field equations (\ref{aux2}) let us write
for the interior and exterior regions the expressions \begin{equation}
\tau^{int}\left(r\right)=-\left(1-\frac{b^{int}}{r}\right)\frac{\Phi^{\prime int}}{r}-\Lambda^{int}\end{equation}
\begin{equation}
\tau^{ext}\left(r\right)=-\left(1-\frac{b^{ext}}{r}\right)\frac{\Phi^{\prime ext}}{r}-\Lambda^{ext}.\end{equation}

\bigskip

Since we assumed that interior static observers feel no tidal forces,
$\Phi^{\prime int}\left(a\right)=0$, we obtain

\begin{equation}
\tau^{int}\left(r\right)=-\Lambda^{int}\end{equation}
\begin{equation}
\tau^{ext}\left(r\right)=-\left(1-\frac{b^{ext}}{r}\right)\frac{\Phi^{\prime ext}}{r}-\Lambda^{ext}\end{equation}

\bigskip

Using equation (\ref{aux9}) for $\Phi^{\prime ext}$ and the tangential
pressure given by (\ref{aux10}), we have

\begin{equation}
\tau^{ext}\left(a\right)=-\frac{\mathcal{P}}{a}\sqrt{-M+\frac{a^{2}}{l^{2}}}-\Lambda^{ext}\label{presion radial}\end{equation}

\bigskip

So, this last equation gives a relation between the radial tension
at the surface and the tangential pressure of the thin-shell. 

\bigskip

\section{Specific Wormhole Solutions }

It is possible to define various functions that represent wormholes.
In general, these solutions could have thin-shells or simply boundary
surfaces. In any case, we will assume, from now on, that $\Phi'^{int}=0$
in order to permit the traversability of the wormhole.

\subsection{Junction with $\mathcal{P}=0$ (Boundary Surface)}

The exterior solution is vacuum with a negative cosmological constant,
so we have $\tau_{ext}=0$ and $\Lambda_{ext}=-\frac{1}{l^{2}}<0$.
If we consider the boundary surface case, i. e. $\mathcal{P}=0$,
equation (\ref{presion radial}) gives \begin{equation}
\Lambda^{ext}=0.\end{equation}

This fact shows that there is no wormhole solution with $\mathcal{P}=0$
in universes with a negative cosmological constant (i.e. we can not
have a BTZ solution outside).

\subsection{Junction with $\mathcal{P}\neq0$ (Thin-Shell)}

Again we have the conditions $\tau_{ext}=0$ and $\Lambda_{ext}<0$
but now we will consider a thin-shell, i. e. $\mathcal{P}\neq0$.
Now. equation (\ref{presion radial}) gives\begin{equation}
\frac{\mathcal{P}}{a}\sqrt{-M+\frac{a^{2}}{l^{2}}}=-\Lambda^{ext}=\frac{1}{l^{2}}\end{equation}

and the form function given by (\ref{aux12}), is now \begin{equation}
b\left(a\right)=\left(1+M\right)a-\frac{a^{3}}{l^{2}}\label{aux19}\end{equation}

and then, the wormhole mass (\ref{masaagujero}) is\begin{equation}
M=\frac{b\left(a\right)}{a}+\frac{a^{2}}{l^{2}}-1.\end{equation}

Is is clear that tha mass associated with the wormhole is zero when
$b\left(a\right)=a-\frac{a^{3}}{l^{2}}$, it is positive when $b\left(a\right)>a-\frac{a^{3}}{l^{2}}$
and it is negative if $b\left(a\right)<a-\frac{a^{3}}{l^{2}}$.

For the subsequent steps, we wil cosider the limit case $b\left(a\right)=a-\frac{a^{3}}{l^{2}}$.
Choosing the form function we will obtain different wormholes. Here,
we consider only two possible functions.

\bigskip

\begin{enumerate}
\item First, consider the functions \begin{align}
b\left(r\right) & =\left(r_{m}r\right)^{\frac{1}{2}}\\
\Phi\left(r\right) & =\Phi_{o}\nonumber \end{align}

where $r_{m}$ is the throat radius. We have\begin{align}
b^{\prime}\left(r\right) & =\frac{1}{2}\sqrt{\frac{r_{m}}{r}}\\
\Phi^{\prime}\left(r\right) & =0.\nonumber \end{align}

The field equations (\ref{aux14}) to (\ref{aux15}) can be written
as \begin{equation}
\overline{\rho}\left(r\right)\equiv\rho\left(r\right)+\Lambda_{int}=-\frac{1}{4r^{3}}\sqrt{r_{m}r}\end{equation}
\begin{equation}
\overline{\tau}\left(r\right)\equiv\tau\left(r\right)+\Lambda_{int}=0\end{equation}
\begin{equation}
\overline{p}\left(r\right)\equiv p\left(r\right)-\Lambda_{int}=0.\end{equation}

Note that in this case the matter density $\rho$ can be negative,
positive or zero, depending on the value of the internal cosmological
constant $\Lambda_{int}$. The total matter density $\overline{\rho}$
is always negative and correspond to the function shown in the Figure
1. 

\begin{center}
\includegraphics[scale=0.3]{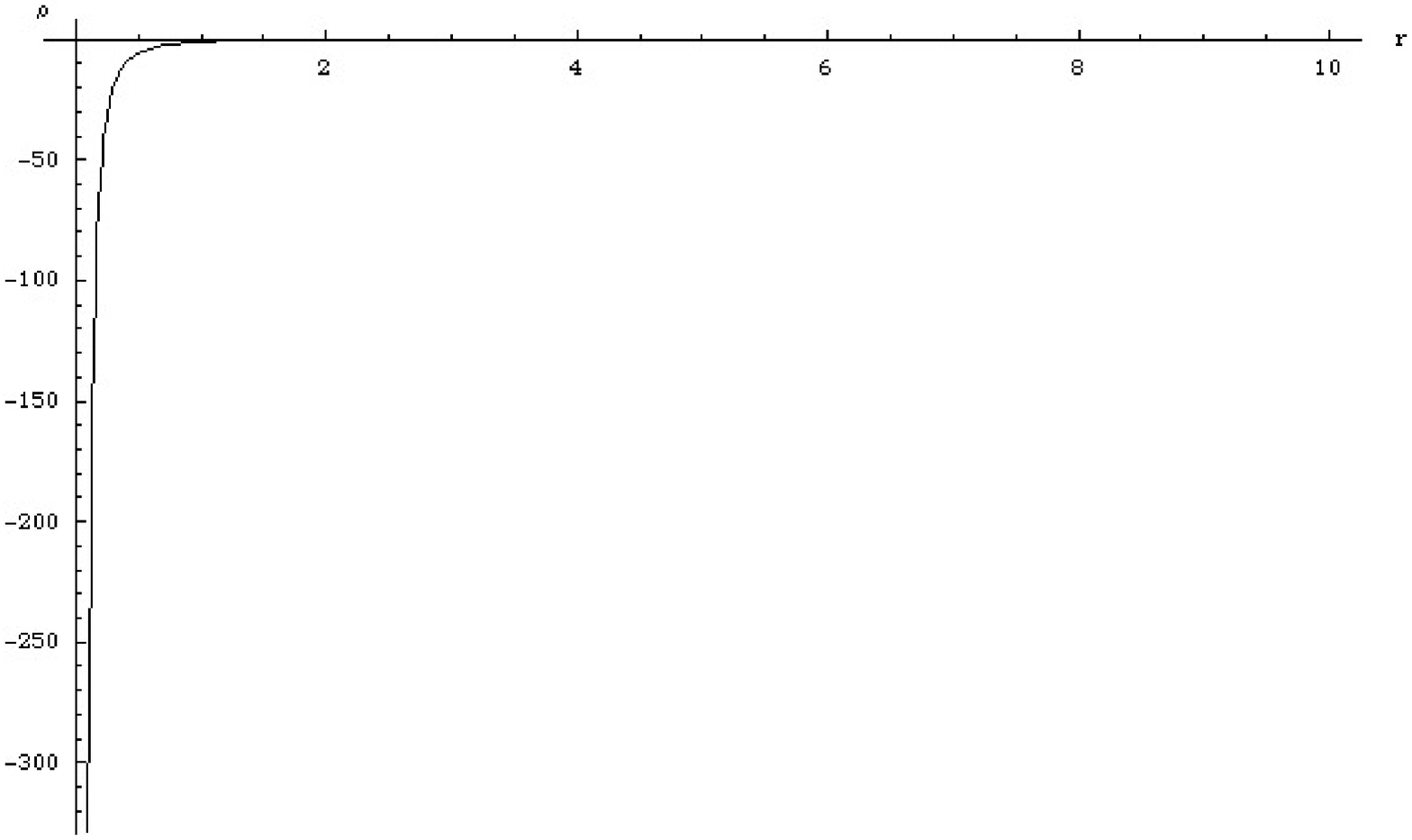}
\par\end{center}

\begin{center}
{\footnotesize Figure 1. Mass density for the first wormhole solution.
Note that this corresponds to exotic matter, and it is distributed
in all the wormhole.}
\par\end{center}{\footnotesize \par}

Equation (\ref{aux19}) gives\begin{equation}
b\left(a\right)=\left(r_{m}a\right)^{\frac{1}{2}}=a-\frac{a^{3}}{l^{2}}\end{equation}
\begin{equation}
\frac{a^{2}}{l^{2}}=1-\left(\frac{r_{m}}{a}\right)^{\frac{1}{2}},\end{equation}
and in order to obtain a wormhole and not a black hole, we must impose
$a>r_{+}$, that gives \begin{equation}
a>\frac{r_{m}}{\left(M-1\right)^{2}}.\end{equation}
Moreover, the constant $\Phi_{o}$ must satisfy $e^{2\Phi\left(a\right)}=-M+\frac{a^{2}}{l^{2}}$
$\,\,$(equation \ref{aux11}). Therefore\begin{equation}
e^{2\Phi_{o}}=-M+\frac{a^{2}}{l^{2}}\end{equation}
Finally, the metric is: in the interior, ($r_{m}\leq r\leq a$),\begin{equation}
ds^{2}=-\left(-M+\frac{a^{2}}{l^{2}}\right)c^{2}dt^{2}+\frac{dr^{2}}{\left(1-\sqrt{\frac{r_{m}}{r}}\right)}+r^{2}d\varphi^{2}\end{equation}
while in the exterior, ($a\leq r\leq\infty$), the metric is\begin{equation}
ds^{2}=-\left(-M+\frac{r^{2}}{l^{2}}\right)dt^{2}+\frac{dr^{2}}{\left(-M+\frac{r^{2}}{l^{2}}\right)}+r^{2}d\varphi^{2}.\end{equation}

\item Our second option of wormhole is \begin{align}
b\left(r\right) & =\frac{r_{m}^{2}}{r}\\
\Phi\left(r\right) & =\Phi_{o}\nonumber \end{align}

with $r_{m}$ the throat radius. Now, we have\begin{align}
b^{\prime}\left(r\right) & =-\frac{r_{m}^{2}}{r^{2}}\\
\Phi^{\prime}\left(r\right) & =0.\nonumber \end{align}

The field equations are now given by \begin{equation}
\overline{\rho}\left(r\right)\equiv\rho\left(r\right)+\Lambda_{int}=\frac{1}{2r^{3}}\left[-\frac{r_{m}^{2}}{r^{2}}r-\frac{r_{m}^{2}}{r}\right]=-\frac{r_{m}^{2}}{r^{5}}\end{equation}
\begin{equation}
\overline{\tau}\left(r\right)\equiv\tau\left(r\right)+\Lambda_{int}=0\end{equation}
\begin{equation}
\overline{p}\left(r\right)\equiv p\left(r\right)-\Lambda_{int}=0.\end{equation}

Note that the value of the mass density $\rho$ is negative, positive
or zero depending on the value of the internal cosmological constant
$\Lambda_{int}$, while the total mass density $\overline{\rho}$
is always negative and behaves like is shown in Figure 2.

\begin{center}
\includegraphics[scale=0.3]{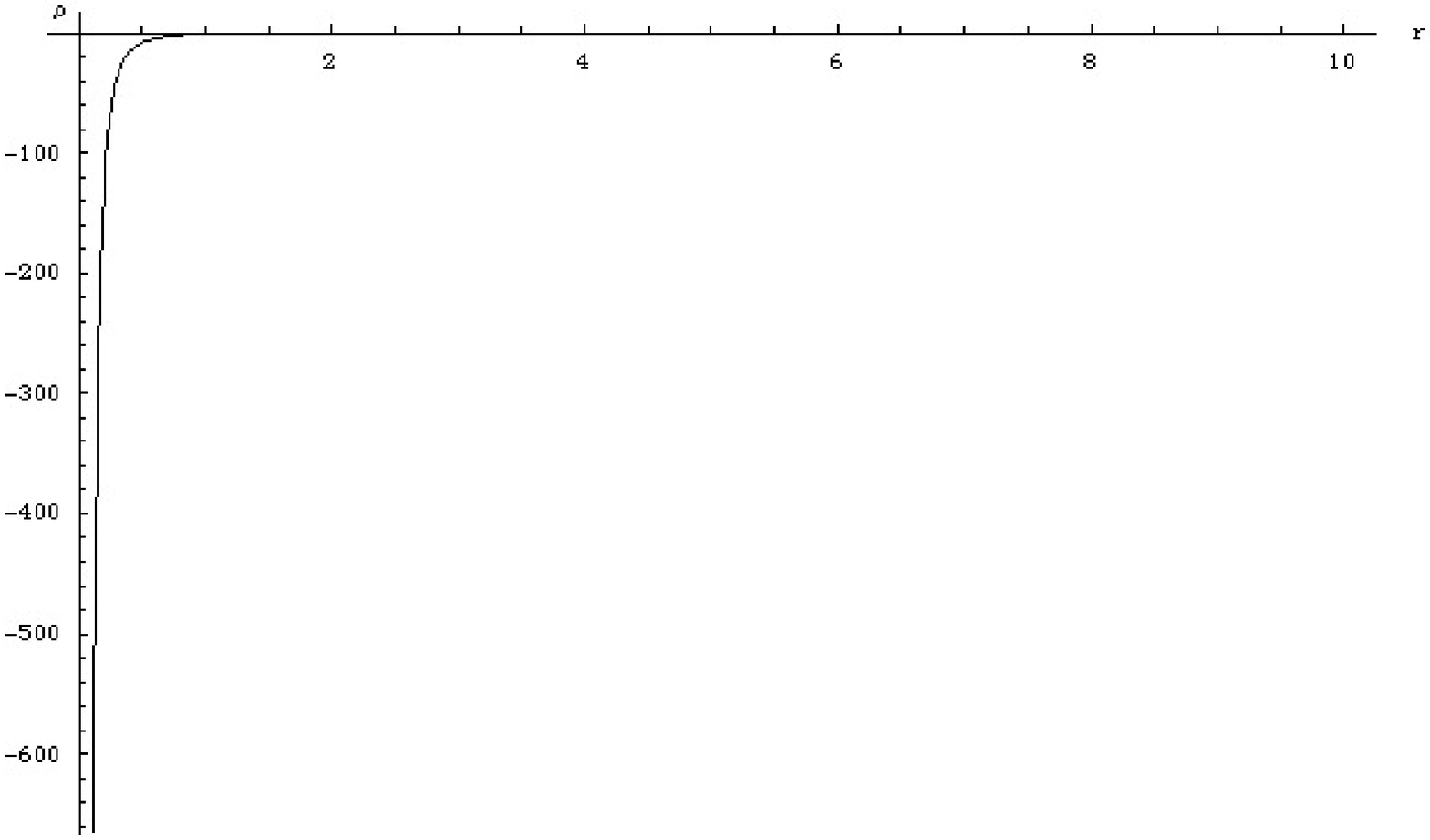}
\par\end{center}

\begin{center}
{\footnotesize Figure 2. Mass density for the second wormhole solution.
Again, this corresponds to exotic matter, and it is distributed in
all the wormhole.}
\par\end{center}{\footnotesize \par}

Using equation (\ref{aux16}) we obtain \begin{equation}
b\left(a\right)=\frac{r_{m}^{2}}{a}=a-\frac{a^{3}}{l^{2}}\end{equation}
\begin{equation}
r_{m}^{2}=a^{2}-\frac{a^{4}}{l^{2}}.\end{equation}

So, the wormhole mouth must be located at\begin{equation}
a^{2}=\frac{l^{2}}{2}\left[1\pm\sqrt{1-4\frac{r_{m}^{2}}{l^{2}}}\right].\end{equation}

To obtain a wormhole solution and not a black hole we must impose
the condition $a>r_{+}$, that gives \begin{equation}
1\pm\sqrt{1-4\frac{r_{m}^{2}}{l^{2}}}>2M.\end{equation}

The constant $\Phi_{o}$ must satisfy $e^{2\Phi\left(a\right)}=-M+\frac{a^{2}}{l^{2}}$
again$\,\,$(equation \ref{aux11}). Then\begin{equation}
e^{2\Phi_{o}}=-M+\frac{a^{2}}{l^{2}}.\end{equation}

Finally, the metric of the wormhole is, in the interior $r_{m}\leq r\leq a$,
\begin{equation}
ds^{2}=-\left(-M+\frac{a^{2}}{l^{2}}\right)c^{2}dt^{2}+\frac{dr^{2}}{\left(1-\frac{r_{m}^{2}}{r^{2}}\right)}+r^{2}d\varphi^{2}\end{equation}

and in the exterior, $a\leq r\leq\infty,$\begin{equation}
ds^{2}=-\left(-M+\frac{r^{2}}{l^{2}}\right)dt^{2}+\frac{dr^{2}}{\left(-M+\frac{r^{2}}{l^{2}}\right)}+r^{2}d\varphi^{2}.\end{equation}

\end{enumerate}
\bigskip

\section{Conclusion}

In this paper we have consider the usual method for construction of
wormholes, by joining two spacetimes in the formalism of ($2+1$)
dimensional gravity with a negative cosmological constant. In the
internal region we impose a appropiate geometry to obtain a traversable
wormhole, while, in the exterior region we use a BTZ black hole solution.
In this way we obtain two specific representing traversable wormholes.
It is also shown that both solutions need of some exotic matter to
exist.

$\allowbreak$
\end{document}